\documentclass[%
 reprint,
 amsmath,amssymb,
 aps,
]{revtex4-1}

\usepackage{graphicx}
\usepackage{dcolumn}
\usepackage{bm}

\usepackage{color}					

\newcommand{\TT}{\mathsf{T}}
\newcommand{\NN}{\mathsf{N}}
\newcommand{\LL}{\mathsf{L}}

\usepackage [english]{babel}
\usepackage [english = american]{csquotes}
\MakeOuterQuote{"}

\begin{document}


\title{Cellular automata as convolutional neural networks}

\author{William Gilpin}
 \altaffiliation[Also at ]{Quantitative Biology Initiative, Harvard University}
 \email{wgilpin@stanford.edu}
\affiliation{%
Department of Applied Physics, Stanford University\\
}%

\date{\today}

\begin{abstract}
Deep learning techniques have recently demonstrated broad success in predicting complex dynamical systems ranging from turbulence to human speech, motivating broader questions about how neural networks encode and represent dynamical rules. We explore this problem in the context of cellular automata (CA), simple dynamical systems that are intrinsically discrete and thus difficult to analyze using standard tools from dynamical systems theory. We show that any CA may readily be represented using a convolutional neural network with a network-in-network architecture. This motivates our development of a general convolutional multilayer perceptron architecture, which we find can learn the dynamical rules for arbitrary CA when given videos of the CA as training data. In the limit of large network widths, we find that training dynamics are nearly identical across replicates, and that common patterns emerge in the structure of networks trained on different CA rulesets. We train ensembles of networks on randomly-sampled CA, and we probe how the trained networks internally represent the CA rules using an information-theoretic technique based on distributions of layer activation patterns. We find that CA with simpler rule tables produce trained networks with hierarchical structure and layer specialization, while more complex CA produce shallower representations---illustrating how the underlying complexity of the CA's rules influences the specificity of these internal representations. Our results suggest how the entropy of a physical process can affect its representation when learned by neural networks.
\end{abstract}

\maketitle

\section{Introduction}

Recent studies have demonstrated the surprising ability of deep neural networks to learn predictive representations of dynamical systems \cite{zdeborova2017machine,pathak2018model,carrasquilla2017machine,van2017learning,torlai2018neural}. For example, certain types of recurrent neural networks, when trained on short-timescale samples of a high-dimensional chaotic process, can learn transition operators for that process that rival traditional simulation techniques \cite{pathak2018model,jaeger2004harnessing,bar2018data}. More broadly, neural networks can learn and predict general features of dynamical systems---ranging from turbulent energy spectra \cite{kutz2017deep}, to Hamiltonian ground states \cite{carleo2017solving,torlai2016learning}, to topological invariants \cite{zhang2018machine}. Such successes mirror well-known findings in applied domains \cite{lecun2015deep}, which have convincingly demonstrated that neural networks may not only represent, but also learn, generators for processes ranging from speech generation \cite{van2016wavenet} to video prediction \cite{mathieu2015deep}. However, open questions remain about how the underlying structure of a physical process affects its representation by a neural network trained using standard optimization techniques.

We aim to study such questions in the context of cellular automata (CA), among the simplest dynamical systems due to the underlying discreteness of both their domain and the dynamical variables that they model. The most widely-known CA is Conway's Game of Life, which consists of an infinite square grid of sites 
("cells") that can only take on a value of zero ("dead") or one ("alive"). Starting from an initial binary pattern, each cell is synchronously updated based on its current state, as well as its current number of living and non-living neighbors. Despite its simple dynamical rules, the Game of Life has been found to exhibit remarkable properties ranging from self-replication to Turing universality \cite{adamatzky2010game}. Such versatility offers a vignette of broader questions in CA research, because many CA offer minimal examples of complexity emerging from apparent simplicity \cite{wolfram1983statistical,langton1990computation,feldman2008organization,fredkin1990informational,adamatzky2012collision}. For this reason, CA have  previously been natural candidates for evaluating the expressivity and capability of machine learning techniques such as genetic algorithms \cite{mitchell1996evolving,mitchell1993revisiting}.

Here, we show that deep convolutional neural networks are capable of representing arbitrary cellular automata, and we demonstrate an example network architecture that smoothly and repeatably learns an arbitrary CA using standard loss gradient-based training. Our approach takes advantage of the "mean field limit" for large networks \cite{nguyen2019mean,neal2012bayesian,chen2018dynamical}, for which we find that trained networks express a universal sparse representation of CA based on depthwise consolidation of similar inputs. The effective depth of this representation, however, depends on the entropy of the CA's underlying rules.

\section{Equivalence between cellular automata and convolutional neural networks}

{\it Cellular automata.} We define a CA as a dynamical system with $M$ possible states, which updates its value based on its current value and $D$ other cells---usually its immediate neighbors in a square lattice. There are $M^D$ possible unique $M$-ary input strings to a CA function, which we individually refer to as $\sigma$. A cellular automaton implements an operator $\mathcal G(\sigma)$ that is fully specified by a list of transition rules $\sigma \rightarrow m$, $m \in 0, 1, ..., M-1$, and there are $M^{M^D}$ possible unique $\mathcal G(\sigma)$, each implementing a different ruleset. For the Game of Life, $M=2, D=9$, and so $\mathcal G(\sigma)$ is a Boolean function that maps each of the $2^9=512$ possible $9$-bit input strings to a single bit. A defining feature of CA is the locality of dynamical update rule, which ensures that the rule domain is small; the size of $D$ thus sets an upper bound on the rate at which information propagates across space. 

{\it Convolutional neural networks.} We define a convolutional neural network as a function that takes as an input a multichannel image, to which it applies a series of local convolutions via a trainable "kernel". The same kernel is applied to all pixels in the image, and each convolutional layer consolidates information within a fixed local radius of each pixel in the input image  \cite{lecun2015deep}. Many standard convolutional architectures include "pooling" layers, which downsample the previous layer and thereby consolidate local information across progressively larger spatial scales; however, all CNN discussed in this paper do not include downsampling steps, and thus preserve the full dimensionality of the input image.

{\it Cellular automata as recurrent mlpconv networks.} The primary analogy between cellular automata and traditional convolutional neural networks arises from (1) the locality of the dynamics, and (2) simultaneous temporal updating of all spatial points. Because neural networks can, in principle, act as universal function approximators \cite{cybenko1989approximation}, a sufficiently complex neural network architecture can be used to fully approximate each rule $\sigma \rightarrow m$ that comprises the CA function $\mathcal G(\sigma)$. This single-neighborhood operator can then be implemented as a convolutional operator as part of a CNN, allowing it to be applied synchronously to all pixel neighborhoods in an input image. 

Representing a CA with a CNN thus requires two steps: feature extraction in order to identify each of the $M^D$ input cases describing each neighborhood, followed by association of each neighborhood with an appropriate output pixel. In the appendix, we show explicitly how to represent {\it any} CA using a single convolutional layer, followed by repeated $1\times1$ convolutional layers. The appropriate weights can be found analytically using analysis of the CA itself, rather than via algorithmic training on input data. In fact, we find that many representations are possible; we show that one possible approach defines a shallow network that uniquely matches each of the $M^D$ input $\sigma$ against a template, while another approach treats layers of the network like levels in a tree search that iteratively narrows down each input $\sigma$ to the desired output $m$. A key aspect of our approach is our usage of only one non-unity convolutional layer (with size $3 \times 3$ for the case of the Game of Life), which serves as the first hidden layer in the network. The receptive field of these convolutional neurons is equivalent to the neighborhood $D$ of the CA. All subsequent layers consist of $1 \times 1$ convolutions, which do not consolidate any additional neighbor information.

Our use of $1\times1$ convolutions to implement the logic of the CA rule table is inspired by recent work showing that such layers can greatly increase network expressivity at low computational cost \cite{szegedy2015going}. Moreover, because CA are explicitly local, the network requires no pooling layers---making the network the equivalent of fitting a small, convolutional multilayer perceptron or "mlpconv" to the CA \cite{lin2013network,szegedy2015going}. Our general approach is comparable to previous uses of deep convolutional networks to parallelize simple operations such as binary arithmetic \cite{kaiser2016neural}, and it differs from efforts using less-common network types with sigma-pi units, in which individual input bits can gate one another \cite{wulff1993learning}. 

\begin{figure*}
{
\centering
\includegraphics[width=.8\linewidth]{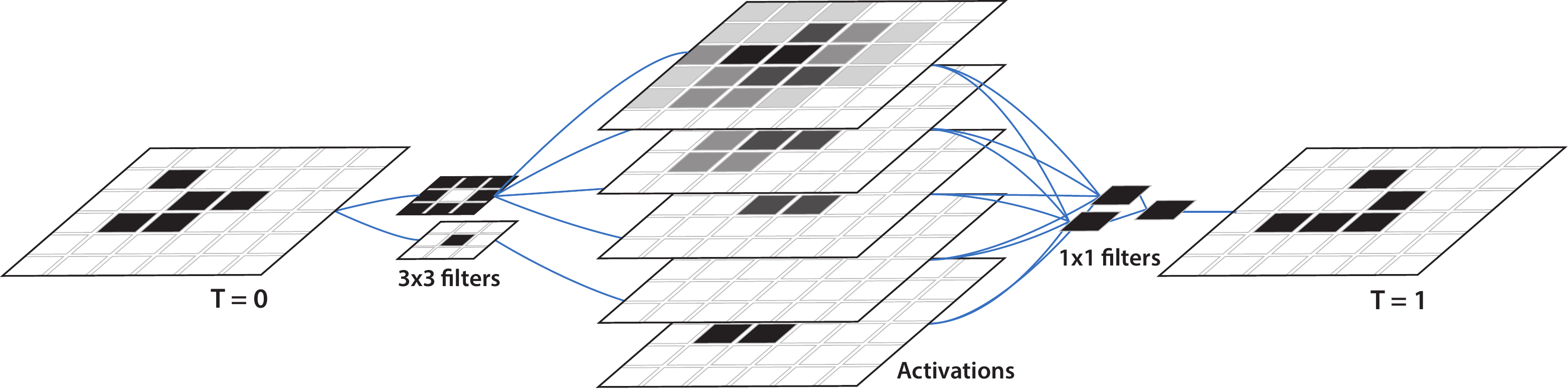}
\caption{
{\bf Conway's Game of Life as a convolutional neural network.}  Two convolutional filters identify the value of the center pixel and count the number of neighbors. These features are then scored and summed to generate a prediction for the system at the next timepoint.
}
\label{model}
}
\end{figure*}

Figure \ref{model} shows an example analytical mlpconv representation of the Game of Life, in which the two salient features for determining the CA evolution (the center pixel value and the number of neighbors) are extracted via an initial $3\times3$ convolution, the results of which are passed to additional $1\times1$ convolutional layers in order to generate a final output prediction (exact weights are given in Supplementary Material). The number of separate convolutions (four with the neighbor filter with different biases, and one with the identity filter) is affected by our choice of ReLU activations (the current best practice for deep convolutional networks) instead of traditional neurons with saturating nonlinearities \cite{nair2010rectified}. Many alternative and equivalent representations may be defined, underscoring the expressivity of multilayer perceptrons when representing simple functions like CA.

\section{A general network architecture for learning arbitrary cellular automata}

Having proven that arbitrary cellular automata may be analytically represented by convolutional perceptrons with finite layers and units, we next ask whether automated training of neural networks on time series of cellular automata images is sufficient to learn their rules. We investigate this process by training ensembles of convolutional neural networks on random random images and random CA rulesets. We start by defining a CA as an explicit mapping between each of $2^9=512$ possible $3\times3$ pixel groups in a binary image, and a single output pixel value. We then apply this map to an ensemble of random binary images (the training data), in order to produce a new output binary image set (the training labels). Here, we use large enough images ($10\times10$ pixels) and training data batches ($500$ images) to ensure that the training data contains at least one instance of each rule. On average, each image contains an equal number of black and white pixels; for sufficiently large images this ensures that each of the $512$ input states is equally probable. We note that, in principle, training the network will proceed much faster if the network is shown an example of only one rule at a time. However, such a process causes the network structure to depend strongly on the order in which individual rules were shown, whereas presenting all input cases simultaneously forces the network to learn internal rule representations based on their relative importance for maximizing accuracy.

\begin{figure*}
{
\centering
\includegraphics[width=.8\linewidth]{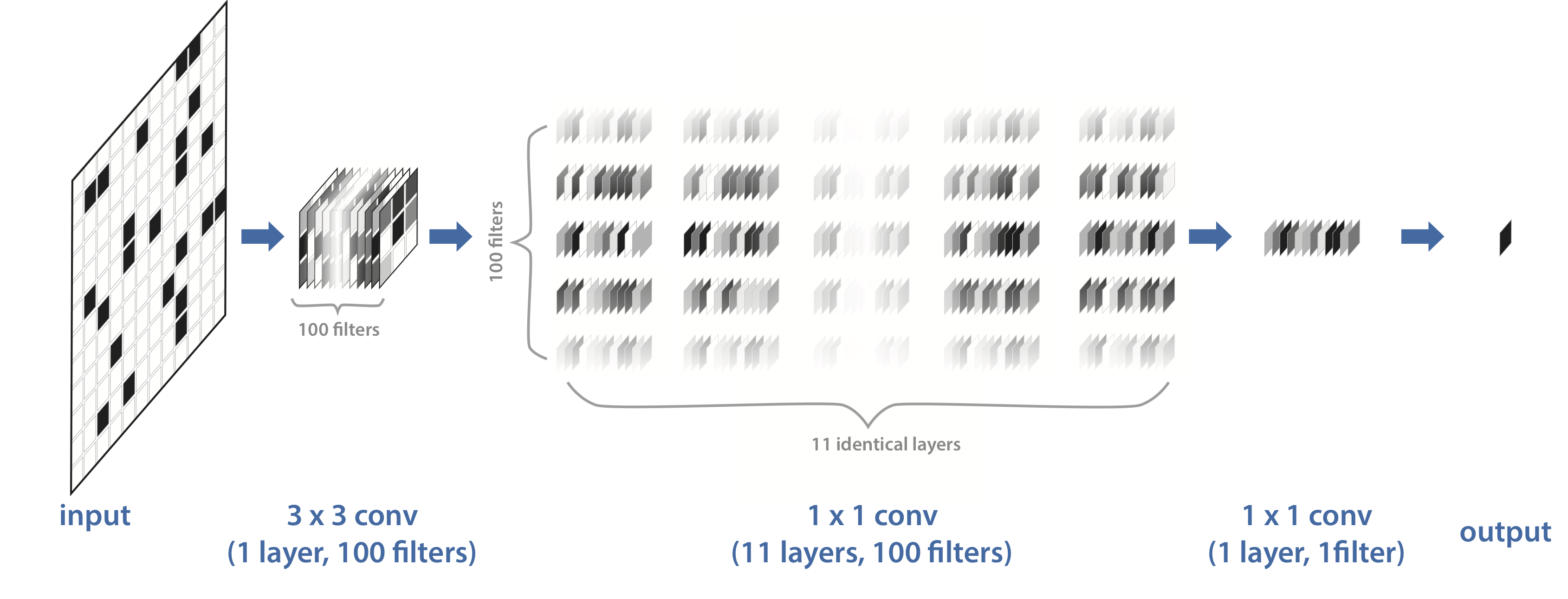}
\caption{
{\bf Architecture of a trainable convolutional neural network for learning cellular automata.} A schematic of the mlpconv network trained on binary cellular automata. Dimensions, where not marked, are determined by the dimensionality of the previous layer.
}
\label{mlpconv}
}
\end{figure*}

{\it Network architecture and training parameters.} Figure \ref{mlpconv} shows the network used in our training experiments. Our network consists of a basic mlpconv architecture corresponding to a single $3\times3$ convolutional layer, followed by a variable number of $1\times1$ convolutional layers \cite{lin2013network}. No pooling layers are used, and the parameters in the $3\times3$ and $1\times1$ layers are trained together. The final hidden layer consists of a weighted summation, which generates the predicted value for the next state of a lattice site. Empirically, including final "prediction" layer with softmax classifier accelerates training on binary CA by reducing the dependence of convergence on initial neuron weights; however we omit this step here in order to allow the same architecture to readily be generalized for CA with $M>2$. Our network may thus be considered a fully convolutional linear committee machine.

We trained our networks using the Adam optimizer with an L2 norm loss function, with hyperparameters (learning rate, initial weights, etc) optimized via a grid search (see Appendix for all hyperparameters). Because generating new training data is computationally inexpensive, for each stage of hyper parameter tuning, a new, unseen validation dataset was generated. Additionally, validation was performed using randomly-chosen, unseen CA rulesets in order to ensure that network hyperparameters were not tuned to specific CA rulesets. During training, a second validation dataset $20\%$ of the size of the training data was generated from the same CA ruleset. Training was stopped when the network prediction accuracy reached 100\% on this secondary validation dataset, after rounding predictions to the nearest integer. The loss used to compute gradients for the optimizer was not rounded. The final, trained networks were then applied to a new dataset of unseen test data (equal in size to five batches of training data).

We found that training successfully converged for all CA rulesets studied, and we note that our explicit use of a convolutional network architecture simplifies learning of the full rule table. Because we are primarily interested in using CNN as a way to study internal representations of CA rulesets, we emphasize that $100\%$ performance on the second validation dataset a condition of stopping training. As a result, all trained networks had identical performance; however, the duration and dynamics of training varied considerably by CA ruleset (discussed below). Regardless of whether weight-based regularization was used during training, we found that performance on the unseen test data was within $\sim 0.3\%$ of the training data for all networks studied (after outputs are rounded, performance reaches $100\%$, as expected). We caution, however, that this equal train-test performance should not be interpreted as a measure of generalizability, as would be the case for CNN used to classify images, etc. \cite{goodfellow2016deep}. Rather, because a CA only has $M^D$ possible input-output pairs (rather than an unlimited space of inputs), this result simply demonstrates that training was stopped at a point where the model had encountered and learned all inputs. In fact, we note that it would be impossible to train a network to represent an arbitrary CA without being exposed to all of its inputs: since an arbitrary CA can send any given input $\sigma$ to any given output $m$, there is no way for a network to predict the output for an symbol without having encountered it previously. However, we note that a network could, in principle, encode a prior expectation for an unseen input symbol $\sigma$, if it was trained primarily on CA of a certain type.

In a previous work that used a one-layer network to learn the rules of a chaotic CA, it was found that training without weight-sharing prevents full learning, because different spatial regions on the system's attractor have different dynamical complexity \cite{wulff1993learning}. In the results below, we deliberately use very large networks with 12 hidden layers---one $3\times3$ convolutional layer, followed by eleven $1\times1$ convolutional layers, all with 100 neurons per layer. These large networks ensure that the network can represent the CA ruleset in as shallow or deep a manner as it finds---and we expect and observe that many fewer neurons per layer are used than are available.

{\it Training dynamics of networks.} Consistent with prior reports that large networks approach a "mean field" limit \cite{neal2012bayesian,chen2018dynamical,sompolinsky1988chaos}, we find that training is highly repeatable for the large networks that we study, even when different training data is used, different CA rules are learned, or the hyperparameters are altered slightly from their optimal values (although this extends the duration of training). We also find that doubling the depth and width of our networks does not qualitatively affect our results, consistent with a large-network limit. Additionally, we trained alternative networks using a different optimizer (vanilla stochastic gradient descent) and loss function (cross-entropy loss), and found nearly identical internal structure in the trained networks (as discussed below); however, the form of the loss curves during training was more concave for such networks. See the supplementary material for further details of networks and training.

Figure \ref{training}A shows the results of training a single network on the Game of Life, and then applying the trained network to the "glider," a known soliton-like solution to the Game. During the early stages of the training, the activations appear random and intermittent. As training proceeds, the network adjusts to the scale of output values generated by the input data, and then begins to learn clusters of related rules---leading to tightening of the output image and trimming of spurious activation patterns. 

\section{Analysis of trained networks}

We next consider the relevance of our training observations to the general properties of binary cellular automata. Intuition would suggest that certain sets of CA rules are intrinsically easier to learn, regardless of $M$ and $D$; for example, a null CA that sends every input to zero in a single timestep requires a trivial network structure, while the Game of Life should require a structure like Figure \ref{model} that can identify each possible neighborhood count. We thus repeat the training data generation and CA network training process described above, except this time we sample CA at random from the  $2^{2^9} \approx 10^{154}$ possible rulesets for binary CA. The complexity of the dynamics produced by a given rule are generally difficult to ascertain {\it a priori}, and typical efforts to systematically investigate the full CA rule space have focused on comparative simulations of different rules \cite{wolfram1983statistical,langton1990computation}. For example, the Game of Life is a member of a unique set of "Class IV" CA capable of both chaotic and regular dynamics depending on their initial state; membership in this class has been hypothesized to be a prerequisite to supporting computational universality \cite{wolfram1983statistical,adamatzky2010game}. General prediction of dynamical class is an ongoing question in the CA literature \cite{mitchell1996evolving}, however, there is a known, approximate relationship between the complexity of simulated dynamics, and the relative fraction $\lambda$ of transitions to zero and one among the full set of $512$ possible input cases: $\lambda=0$ and $\lambda=1$ correspond to null CA, whereas $\lambda=0.5$ corresponds to CA that sends equal numbers of input cases to $0$ and $1$ \cite{langton1990computation}. This captures the general intuition that CA typically display richer dynamics when they have a broader range of output symbols \cite{feldman2008organization,adamatzky2012collision}. Here, instead of using $\lambda$ directly, we parametrize the space of CA equivalently using the effective "rule entropy," $\mathcal H_{ca}$. We define $\mathcal H_{ca}$ by starting from a maximum-entropy image with a uniform distribution of input symbols ($p_\sigma \approx 1/M^D$ for all $\sigma$), to which we then apply the CA rule once and then record the new distribution of input cases, $p_\sigma'$. The residual  Shannon entropy $\mathcal{H}_{ca} \equiv -\sum_\sigma p'_\sigma \log_2 p'_\sigma$ provides a measure of the degree to which the CA rules compress the space of available states. $\mathcal H_{ca}(\lambda)$ monotonically increases from $\mathcal H_{ca}(0) = 0$ until it reaches a global maximum at $\mathcal H_{ca}(1/2) = 9$, after which it symmetrically decreases back to $\mathcal H_{ca}(1) = 0$.

\begin{figure*}
{
\centering
\includegraphics[width=.8\linewidth]{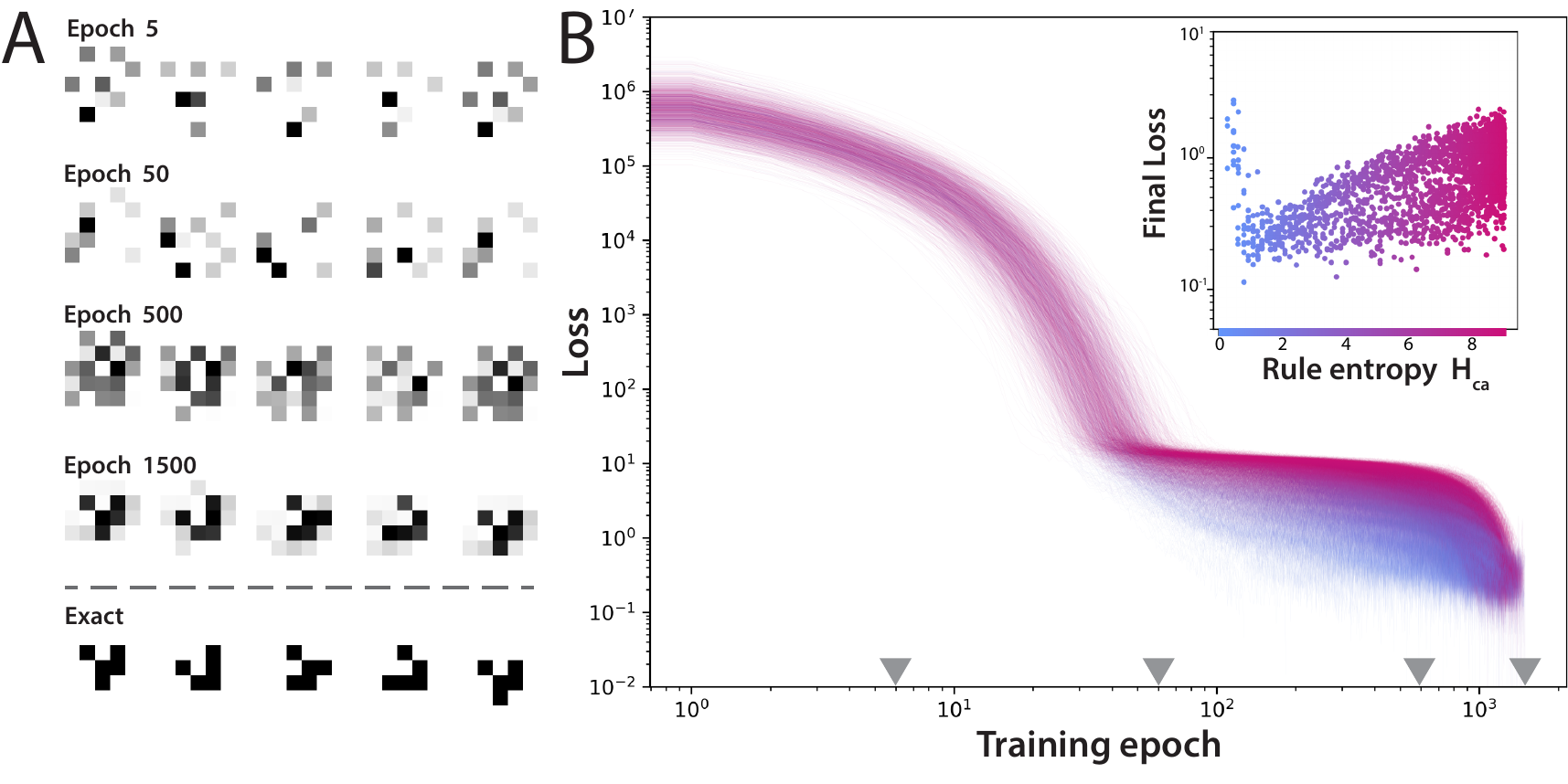}
\caption{
{\bf Training 2560 convolutional neural networks on random cellular automata.} (A) A network trained on the Game of Life for different durations, and then applied to images of each stage of the "glider" solution. (B) The loss versus time during training, colored by the rule entropy $\mathcal H_{ca}$. Groups of 512 related cellular automata were generated by iteratively choosing random $\sigma \rightarrow 0$ rules from the $512$ possible input configurations, and setting those sites to $\sigma \rightarrow 1$. $5$ replicates were performed. Loss values represent the sum over the batch; values of $10$ or smaller imply that only small rounding errors were present at the end of training. The entropy of the resulting rule table is characteristic of the CA, and it is indicated by $\mathcal H_{ca} =0$ (blue, minimum entropy CA) to $\mathcal H_{ca} =9$ (magenta, maximum entropy CA). (Inset) The final loss for each network at the end of training, shown as a function of $\mathcal H_{ca}$.
}
\label{training}
}
\end{figure*}

Figure \ref{training}B shows the result of training $2560$ randomly-sampled CA with different values of $\mathcal H_{ca}$. Ensembles of $512$ related cellular automata were generated by randomly selecting single symbols in the input space to transition to $1$ (starting with the null case $\sigma \rightarrow 0$ for all $\sigma$), one at a time, until reaching the case $\sigma \rightarrow 1$ for all $\sigma$. This "table walk" sampling approach \cite{langton1990computation} was then replicated $5$ times for different starting conditions. 

We observe that the initial $10-100$ training epochs are universal across $\mathcal H_{ca}$. Detailed analysis of the activation patterns across the network (Supplementary material) suggests that this transient corresponds to initialization, wherein the network learns the scale and bounds of the input data. Recent studies of networks trained on real-world data suggest that this initialization period consists of the network finding an optimal representation of the input data \cite{shwartz2017opening}. During the next stage of training, the network begins to learn specific rules: the number of neurons activated in each layer begins to decrease, as the network becomes more selective regarding which inputs provoke non-zero network outputs (see supplementary material). Because $\mathcal H_{ca}$ determines the sparsity of the rule table---and thus the degree to which the rules may be compressed---$\mathcal H_{ca}$ strongly affects the dynamics of this phase of training, with simpler CA learning faster and shallower representations of the rule table, resulting in smaller final loss values (Figure \ref{training}B, inset). This behavior confirms general intuition that more complicated CA rules require more precise representations, making them harder to learn.

A key feature of using large networks to fit simple functions like CA is strong repeatability of training across different initializations and CA rulesets. In the appendix, we reproduce all results shown in the main text using networks with different sizes and depths, and even a different optimizer, loss function, and other hyperparameters, and we report nearly identical results (for both training and test data) as those found using our network architecture described above. On both the training data and test data, we find similar universal training curves that depend on $\mathcal H_{ca}$, as well as distributions of activation patterns. This universality is not observed in "narrow" networks with fewer neurons per layer, for which training proceeds as a series of plateaus in the loss punctuated by large drops when the stochastic optimizer happens upon new rules. In this limit, randomly-chosen CA rulesets will not consistently result in training successfully finding all correct rules and terminating. Moreover, small networks that do terminate do not display apparent patterns when their internal structure is analyzed using the approaches described below---consistent with a random search. Similar loss dynamics have previously been observed when CA are learned using genetic algorithms, in which the loss function remains mostly flat, punctuated by occasional leaps when a mutant encounters a new rule \cite{mitchell1996evolving}. For gradient-based training, similar kinetic trapping occurs in the vicinity of shallow minima or saddle points \cite{saad1995line,dauphin2014identifying}, but these effects are reduced in larger networks such as those used here.

\section{Information-theoretic quantification of activations.} 

That training thousands of arbitrary CA yields extremely similar training dynamics suggests that deep networks trained using gradient optimizers learn a universal approach to approximating simple functions like CA. This motivates us to next investigate how exactly the trained networks represent the underlying CA rule table---do the networks simply match entire input patterns, or do they learn consolidated features such as neighbor counts? Because the intrinsic entropy of the CA rule table affects training, we reason that the entropy of activated representations at each layer is a natural heuristic for analyzing the internal states of the network. We thus define a binary measure of activity for each neuron in a fully-trained network: when the network encounters a given input $\sigma$, any neurons that produce a non-zero output are marked as $1$ (or $0$ otherwise), resulting in a new set of binary strings $a(\sigma)$ denoting the rounded activation pattern for each input $\sigma$. For example, in an mlpconv network with only 3 layers, and 3 neurons per layer, an example activation pattern for a specific input $\sigma_1$ could yield $a(\sigma_1) = \{010,000,011\}$, with commas demarcating layers. Our approach constitutes a simplified version of efforts to study deep neural networks by inspecting activation pattern "images" of neurons in downstream layers when specific input images are fed into the network \cite{schoenholz2016deep,erhan2009visualizing,poole2016exponential,chen2018dynamical}. However, for our system binary strings (thresholded activation patterns) are sufficient to characterize the trained networks, due to the finite space of input-output pairs for binary CA, and the large size of our networks; in our investigations, no cases were found in which two different inputs ($\sigma, \sigma'$) produced different unrounded activation patterns, but identical patterns after binarization  ($a(\sigma), a(\sigma')$).

Given the ensemble of input symbols $\sigma\in\{0,1\}^D$, and a network consisting of $L$ layers each containing $N$ neurons, we can define separate symbol spaces representing activations of the entire network $a_\TT(\sigma) \in\{0,1\}^{LN}$; each individual layer, $a_{\LL,i}(\sigma) \in\{0,1\}^{N}$, $i \in [0,L-1]$; and each individual neuron $a_{\NN,ij}(\sigma) \in\{0,1\}$, $i \in [0,L-1]$, $j \in [0,N-1]$. Averaging over test data consisting of an equiprobable ensemble of all $M^D$ unique input cases $\sigma$, we can then calculate the probability $p_{\alpha,k}$ for observing a given unique symbol $a_{k}$ at a level $\alpha \in \{\TT, \LL, \NN\}$ in the network. We quantify the uniformity of each activation symbol distribution $p$ using the entropy $\mathcal H_\alpha = -\sum_k p_{\alpha, k} \log_2 p_{\alpha, k}$, which satisfies $\mathcal{H}_\alpha \leq {\text{dim}{(\alpha)}} $. We condense notation and refer to the activation entropies $\mathcal H_\TT$, $\mathcal{H}_{\LL,i}$, $\mathcal{H}_{\NN,{ij}}$ as the total entropy, the entropy of $i^{th}$ layer, and the entropy of the $j^{th}$ neuron in the $i^{th}$ layer. We note that, in addition to readily quantifying the number of unique activation patterns and their uniformity across input cases, the Shannon entropy naturally discounts zero-entropy "dead neurons," a common artifact of training high-dimensional ReLU networks \cite{nair2010rectified}. Our general analysis approach is related to a recently-developed class of techniques for analyzing trained networks \cite{raghu2017svcca}, in which an ensemble of training data (here, a uniform distribution of $\sigma$) is fed into a trained network in order to generate a new statistical observable (here, $\mathcal{H}$).

We expect and observe that $\langle \mathcal{H}_{\NN,{ij}} \rangle_{ij} < \langle \mathcal{H}_{\LL,i} \rangle_{i} \leq \mathcal{H}_\TT$. Unsurprisingly, the maximum entropy of a single neuron is $\log_2{2}=1$, and all multi-neuron layers generate more than two patterns across the test data. We also observe that $\mathcal{H}_\TT \approx 9$ for all networks trained, suggesting that the overall firing patterns in the network differed for every unique input case---even for trivial rules like $\lambda = 0$ where a network with all zero weights and biases would both correctly represent the rule table, and have identical firing patterns for all inputs ($\mathcal{H}_\TT =0$). This effect directly arises from training using gradient-based methods, for which at least some early layers in the network produce unique activation patterns for each $\sigma$ that are never condensed during later training stages. Accordingly, regularization using a total weight cost or dropout both reduce $\mathcal{H}_\TT$.

\begin{figure*}
{
\centering
\includegraphics[width=.8\linewidth]{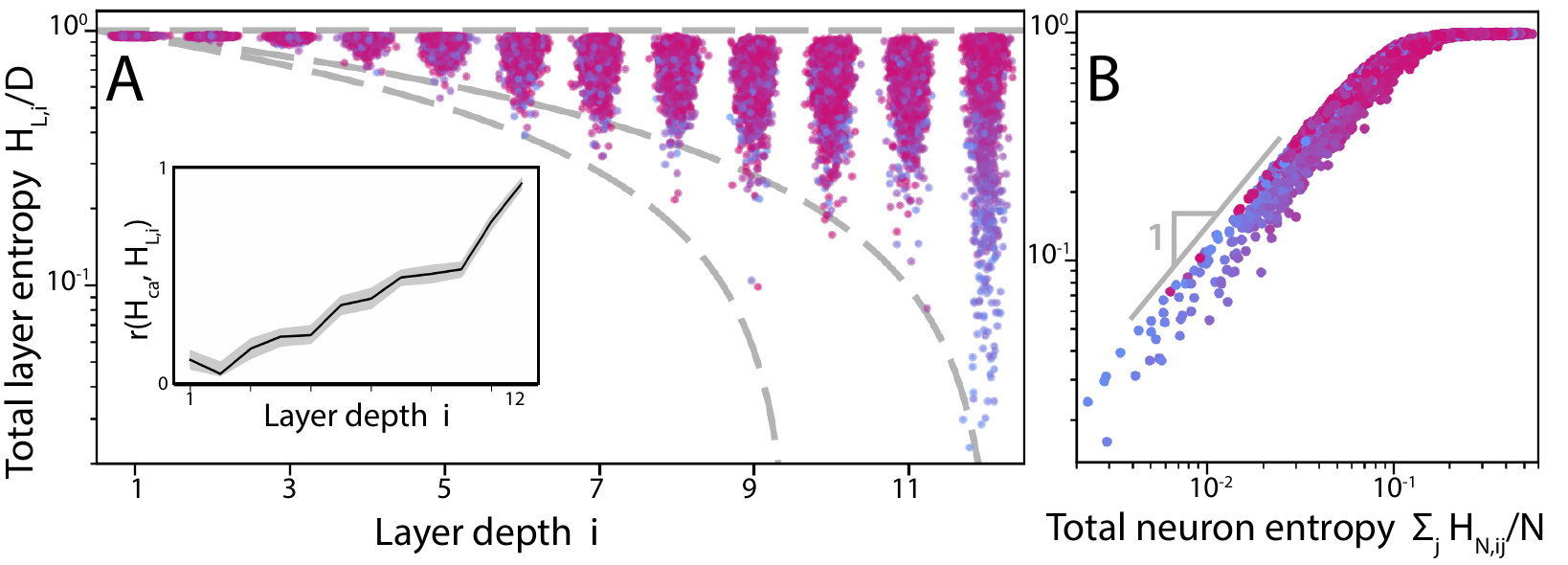}
\caption{
{\bf Internal representations of cellular automata by trained networks.} 
(A) The individual layerwise entropy ($\mathcal{H}_{\LL,i}/D$) for the 2560 networks shown in the previous figure. Noise has been added to the horizontal coordinates (layer index) to facilitate visualization. As in previous figures, coloration corresponds to the entropy $\mathcal H_{ca}$ of the underlying CA. Dashed lines correspond to expected trends for theoretical networks that eliminates $0\%$ of cases in each layer (i.e., a pattern-matching implementation), $45\%$ of cases, and $50\%$ (top to bottom)
(Inset) The Pearson correlation coefficient $r$ between the rule entropy $\mathcal H_{ca}$ and layer entropy $\mathcal{H}_{\LL,i}$. Error range corresponds to bootstrapped 25\% -75\% quantiles.
(B) The normalized layerwise entropy ($\mathcal{H}_{\LL,i}/D$) versus the normalized total layerwise neuron entropy ($\mathcal{H}_{\NN,ij}/N$), with the linear scaling annotated.
}
\label{entropy}
}
\end{figure*}

Comparing $\mathcal{H}_{\LL,i}$ across models and layers demonstrates that early layers in the network tend to generate a broad set of activation patterns that closely follow the uniform input symbol distribution (Figure \ref{entropy}A). These early layers in the network thus remain saturated at $\mathcal{H}_{\LL,i} = \mathcal{H}_\TT \approx 9$; however in deeper layers progressively lower entropies are observed, consistent with fewer unique activation patterns (and a less uniform distribution across these strings) appearing in later layers. These trends depend strongly on the CA rules (coloration). In the figure, dashed lines allow comparison of $\mathcal{H}_{\LL,i}$ to theoretical predictions for the layerwise entropy for the different types of ways that a CNN can represent the CA. The uppermost dashed curve corresponds to a network that generates a maximum entropy set of $512$ equiprobable activation patterns in each layer. This case corresponds to a "shallow" network that matches each input case to a unique template at each layer. Lower dashed curves correspond to predictions for networks that implement the CA as layerwise search, in which $\sigma$ that map to the same output $m$ are mapped to the same activation pattern at some point before the final layer. This corresponds to a progressive decrease in the number of unique activation patterns in each layer. The two dashed curves shown correspond to theoretical networks that eliminate $45\%$ and $50\%$ of unique activation patterns at each layer. 

We find that higher entropy rules $\mathcal H_{ca}$ (red points) tend to produce shallower networks due to the rule table being less intrinsically compressible; whereas simpler CA (blue points) produce networks with more binary tree-like structure. This relationship has high variance in early layers, making it difficult to visually discern in the panel save for the last layer. However, explicit calculation of the Pearson correlation $r(\mathcal H_{ca},\mathcal H_{\LL,i})$ confirms its presence across all layers of the network, and that it becomes more prominent in deeper layers (Figure \ref{entropy}A, inset). This trend is a consequence of training the network using backpropagation-based techniques, in which loss gradients computed at the final, $L^{th}$ hidden layer are used to update the weights in the previous $(L-1)^{th}$ layer, which are then used to update the $(L-2)^{th}$ layer, and so forth \cite{arora2014provable}. During training, the entropy of the final layer increases continuously until it reaches a plateau determined by the network size and by $\mathcal H_{ca}$. The penultimate layer then increases in entropy until reaching a plateau, and so forth until $\mathcal{H}_\TT = 9$ across all $\sigma$---at which point training stops because the test error will reach zero (training dynamics are further analyzed in the Supplementary Material). This general correlation between CA entropy and network structure is consistent with earlier studies in which networks were trained to label CA rulesets by their dynamical complexity class \cite{gorodkin1993neural}.

The role of $\mathcal H_{ca}$ on internal representation distributions $p_L$ can be further analyzed using Zipf plots of activation pattern $a_k$ frequency versus rank (Supplementary Material): the resulting plots show that the distribution of activation symbols is initially uniform (because the training data has a uniform distribution of $\sigma$), but the distribution becomes progressively narrower and more peaked in later layers. This process occurs more sharply for networks trained on CA with smaller $\mathcal H_{ca}$. 

We next consider how the entropy of our observed layer activation patterns relates to the entropy of the individual neurons $\mathcal{H}_{\NN,{ij}}$ that comprise them; we suspect there is a relation because the individual firing entropies determine the "effective" number of neurons in a layer, $N_{\text{eff}} = 2^{\sum_j \mathcal{H}_{\NN,{ij}}}$. Across all layers, we observe a linear relationship between $\mathcal{H}_{\NN,{ij}}$ and $\mathcal{H}_{\LL,i}$, which saturates when $\mathcal{H}_{\LL,i} \approx \mathcal{H}_\TT$ (Figure \ref{entropy}B). The lower-$\mathcal H_{ca}$ CA lie within the linear portion of this plot, suggesting that variation in activation patterns in this regime results from layers recruiting varying numbers of neurons. Conversely, higher-entropy CA localize in a saturated region where each layer encodes a unique activation pattern for each unique input state, leading to no dependence on the total effective number of neurons. This plot explains our earlier observation that the dynamics of training do not depend on the exact network shape as long as the network has sufficiently many neurons: for low $\mathcal H_{ca}$, layers never saturate, and are free to recruit more neurons until they are able to pattern-match every unique input (at intermediate and large $\mathcal H_{ca}$). A CA with more possible input states (larger $M$ or $D$) would thus require more neurons per layer to enter this large-network limit.

We also consider the degree to which the decrease $\mathcal{H}_{\LL,i}$ vs. $i$ arises from deeper layers becoming "specialized" to specific input features, a common observation for deep neural networks \cite{arora2014provable,lecun2015deep,erhan2009visualizing}. We quantify the layer specialization using the total correlation, a measure of the mutual information between the activation patterns of a layer, and the neurons within that layer: $\mathcal I_i=  \sum_j \mathcal{H}_{\NN,{ij}} - \mathcal{H}_{\LL,i}$. This quantity is minimized ($\mathcal I_i=0$) when the single neuron activations within a layer are independent of one another; conversely, at the maximum value individual neurons only activate jointly in the context of forming a specific layer activation pattern. Plots of $\mathcal I_i$ vs. $i$ (Supplementary material) reveal that during early layers, individual neurons tend to fire independently, consistent with multi-neuron features being unique to each input case. In these early layers, $\mathcal I_i$ is large because the number of possible activation patterns in a single layer of the large network ($2^{100}$) is much larger than the number of input cases ($2^9$). In later layers, however, the correlation begins to decrease, consistent with individual neurons being activated in the context of multiple input cases---indicating that these neurons are associated with features found in multiple input cases, like the states of specific neighbors. Calculation of $r(\mathcal I_i, \mathcal H_{ca})$ confirms that this effect varies with $\mathcal H_{ca}$.

\section{Discussion}
We have shown an analogy between convolutional neural networks and cellular automata, and demonstrated a type of network capable of learning arbitrary binary CA using standard techniques. Our approach uses a simple architecture that applies a single $3 \times 3$ convolutional layer in order to consolidate the neighborhood structure, followed by repeated $1\times1$ convolutions that perform local operations. This architecture is capable of predicting output states using a mixture of shallow pattern-matching and deep layer-wise tree searching. After training an ensemble of networks on a variety of CA, we find that our networks structurally encode generic dynamical features of CA, such as the relative entropy of the rule table. Further work is necessary to determine whether neural networks can more broadly inform efforts to understand the dynamical space of CA, including fundamental efforts to relate a CA's {\it a priori} rules to the its apparent dynamical complexity during simulation \cite{wolfram1983statistical,mitchell1993revisiting,feldman2008organization}---for example, do Class IV and other complex CA impose unique structures upon fitted neural networks, or can neural networks predict their computational complexity given a rule table? These problems and more general studies of dynamical systems will require more sophisticated approaches, such as unsupervised training and generative architectures (such as restricted Boltzmann machines). More broadly, we note that studying the bounded space of CA has motivated our development of general entropy-based approaches to probing trained neural networks. In future work we hope to relate our observations to more general patterns observed in studies of deep networks, such as the information bottleneck \cite{shwartz2017opening}. Such results may inform analysis of open-ended dynamical prediction tasks, such as video prediction, by showing a simple manner in which process complexity manifests as structural motifs. 

\vspace{5mm}
\section{Acknowledgments}

W.G. was supported by the NSF-Simons Center for Mathematical and Statistical Analysis of Biology at Harvard University, from NSF Grant No. DMS-1764269, and from the Harvard FAS Quantitative Biology Initiative. He was also supported by the U.S. Department of Defense through the NDSEG fellowship program, as well as by the Stanford EDGE-STEM fellowship program

\section{Code availability}

Convolutional neural networks were implemented in Python 3.4 using TensorFlow 1.8 \cite{abadi2016}. Source code is available at \url{https://github.com/williamgilpin/convoca}.

\section{Appendix}

\subsection{Representing arbitrary CA with convolutional neural networks}

Here we show explicitly how a standard {\it mlpconv} multilayer perceptron architecture with ReLU activation is capable of representing an arbitrary $M$ state cellular automaton with a finite depth and neuron count \cite{lin2013network}. We provide the following explicit examples primarily as an illustration of the ways in which $1\times1$ convolutions may be used to implement arbitrary CA using a perceptron; we note that real-world networks trained using optimizers will find many other heuristics and representations. We provide the two analytic cases below for concreteness, and to illustrate two important limits: pattern-matching templates for each unique input across the entire network, or using individual layers to eliminate cases until the appropriate output symbol has been identified.

\subsubsection{Pattern-matching the rule table with a shallow network}

An arbitrary M-state cellular automaton can first be converted into a one-hot binary representation. Given an $L\times L$ image, we seek to generate an $L\times L\times M$ stack of binary activation images:
\begin{enumerate}
\item Convolve the input layer with $M$ distinct $1\times1$ convolutional filters with unit weights, and with biases given by $1,0,-1,...-(M-1)$. Now apply ReLU activation
\item Convolve the resulting image with $M$ $1\times1$ convolutional filters with zero biases. Each of the first $(M-1)$ convolutional filters tests a different consecutive pair $[1,-b, 0,...,0]$, $[0,1,-b,0,...,0]$, $[0,0,1,-b, 0, ..., 0]$, $...$, $[0,...,0,1,-b]$, where $b$ is any positive constant $b\geq M/(M-1)$. The last convolutional filter is the identity $[0,...,0,1]$. Now apply ReLU activation again.
\end{enumerate}
This conversion step is not necessary when working with a binary CA. It requires at total of $(1+M)+M^2$ parameters and two layers to produce an activation volume of dimensions $L\times L\times M$.

We now have an $L\times L\times (M-1)$ array corresponding the one-hot encoding of each pixel's value in an $L\times L$ lattice. We now pattern match each of the $M^D$ possible inputs with its corresponding correct output value. We note that the steps we take below represent an upper bound; if the number of quiescent versus active states in the cellular automaton is known in advance ($=\lambda M^D$, where $\lambda$ is Langton's parameter) \cite{langton1990computation}, then the number of patterns to match (and thus total parameters) may be reduced by a factor of $\lambda$, because only the non-quiescent "active" rules that produce non-zero output values need to be matched.
\begin{enumerate}
\item Construct a block of $M^D$ $S\times S\times(M-1)$ convolutional filters, where $S$ corresponds to the neighborhood size of the CA ($S=3$ for a standard CA with a Moore neighborhood). Each of the $M^D$ filters simply corresponds to an image of each possible input state, with entries equalling one for each non-zero site, and large negative values (greater than $D(M-1)$) at each zero site. For cases when $M>2$, the depth of each convolutional kernel allows exact matching of different non-zero values.
\item Assign a bias to each of the $M^D$ filters based on the cellular automaton's rule table. For $S\times S\times(M-1)$ inputs that should map to a non-zero value $q$, assign a bias of $(q-1)-(L-1)$, where $L$ is the number of non-zero sites in the neighborhood $L \leq D(M-1)$. This ensures that only exact matches to the rule will produce positive values under convolution. For inputs that should map to zero, assign any bias $\geq L$, such as $D(M-1)$.
\item Apply the ReLU function.
\end{enumerate}

\subsubsection{Searching the rule table with a deep network}
\label{appendix_deep_ca}

Another way to represent a cellular automaton with a multilayer perceptron constitutes searching a subset of all possible inputs in each layer. This approach requires all input cases $\sigma$ that map to the same output symbol $m$, to also map to the same activation pattern at some layer of the network. This coalescence of different input states can occur at any point in the network before the final layer; here we outline a general approach for constructing maps to the same output symbol using large networks.

\noindent{\bf Assigning input cases to a unique binary strings.} Assume there are $N$ convolutional filters. If there are $M^D$ unique input cases, these filters can be used to generate an $n$-hot encoding of the input states. $n$ should be chosen such that ${N \choose n} \geq M^D$. Here, we assume a binary CA with a Moore neighborhood ($M=2$, $D=9$). If $N=100$ neurons are present in each layer, then a two-hot binary string ($n=2$) is sufficient to uniquely represent every possible input state of a binary Moore CA, using the following steps

\begin{enumerate}
\item The $D$ pixel neighborhood is split into $n$ sub-neighborhoods, with sizes we refer to as $D_1, D_2, .., D_n$. For example, for a the binary Moore CA, we can split the neighborhood into the first $5$ pixels (counted from top-left to the center) and the remaining $4$ pixels (the center pixel to the bottom right corner. Note that the number and dimensionality of these sub-neighborhoods must satisfy the condition: if $Q \equiv M^{D_1} + M^{D_2} + ... + M^{D_n}$, then ${N \choose Q} \geq M^D$.
\item Define $M^{D_1} + M^{D_2} + ... + M^{D_n}$ filters, which match each possible sub-neighborhood. For example, for the neighborhood reading $101000111$ from upper-left to bottom-right, two filters can be defined that will match sub-neighborhoods consisting of the first $5$ bits and the last $4$ bits, using the approach described above for pattern-matching. In this case, these filters would be $1, -100, 1, -100,-100,-100, 0, 0, 0$ with a bias of $-1$, and $0, 0, 0, -100, -100, -100,1,1,1$ with a bias of $-2$.
\item Apply ReLU activation.
\item The resulting activation map will be an $n$-hot binary encoding of the input state, because each unique input case will match the same $n$ filters from the set of $N$, thus creating a unique representation.
\end{enumerate}

\noindent{\bf Assigning input case binary strings to matching output symbols.} At this stage in the network, each input case has been mapped to a unique $N$ digit binary string with exactly $n$ ones within it. Successive $1\times1$ convolutional filters may now be used to combine different inputs into the same activation pattern. As a simple example, if $N=5$ then the possible input cases are $\sigma \in \{10001$, $10010$,$10100$, $11000$, $01001$, $01010$, $01100$, $00101$, $00110$, $00011\}$. Many of these cases can be uniquely matched by applying a filter consisting of three ones, followed by a bias of $b=2$. For example, using the filter $W = (-1,-2,-1,0,-2)$ to perform the operation $h = RELU(W\cdot\sigma+b)$ will result in an output of $1$ for the cases $\{10010,00110\}$ only. To match strings with no overlapping bits, more than two cases must be merged simultaneously. In general, to merge $H$ cases using this approach, two strings must have $H-1$ overlapping bits.

For the case of binary CA with a Moore radius, an example of a network analogous to a simple binary search would consist of filters that reduce the $512$ input cases to $512$ $2$-hot strings (in the first $3\times3$ convolutional layer). Subsequent $1\times1$ convolutions could then map these states to $256$ unique cases, then $128$, and so forth until there are only two unique activation patterns left---the first for input states that map to one, and the second for input states that map to zero. Depending on the $\lambda$ parameter of the CA rule table, the depth (and thus minimum number of layers) to perform this search would be a maximum of $\log_2 512 - 1 = 8$ layers when $\lambda=0.5$ (i.e. when there are equal numbers of ones and zero outputs in the rule table). This case comprises just one example of performing a search using the depth of a network. However, many variations are possible, because coalescence of two input states may occur in any layer. Moreover, while the above examples describe two input states being combined together for each filter in a given layer, it is not difficult to construct alternative filters that can combine more than two states together. We thus expect that there is considerably flexibility in the different ways that a network trained algorithmically can internally represent input states with similar features and similar outputs, but that these different approaches manifest as an overall decrease in the number of unique activation patterns observed across the depth of the network.

\subsubsection{Network representation of the Game of Life}

We note that there are many other ways to implement a CA that are not exactly layerwise depth search, nor a shallow pattern match, depending on the number and type of features being checked at each layer of the network. For example, each of the $D$ pixels in the neighborhood of the CA can be checked with separate convolutional kernels all in the first layer, and then different combinations of these values could be checked in subsequent steps. The shallow network described above represents an extreme case, in which every value of the full input space is explicitly checked in the first layer. This implementation is efficient for many CA, because of the low cost of performing multiple numerical convolutions. However, for CA with large $M$ or $D$, the layer-wise search method may be preferable.

For the Game of Life, we can use knowledge of the structure of a CA in order to design a better implementation. The Game of Life is an outer totalistic CA, meaning that the next state of the system is fully determined by the current value of the center pixel, and the total number of ones and zeros among its immediate neighbors. For this reason, only two unique convolutional filters are needed.

The first filter is the identity,
\begin{table}[htp]
\centering
\begin{tabular} {|ccc|}
\hline
0 & 0 & 0\\
0 & 1 & 0\\
0 & 0 & 0\\
\hline
\end{tabular}
\end{table}
which is applied with bias $0$.

The second filter is the neighbor counting filter
\begin{table}[htp]
\centering
\begin{tabular} {|ccc|}
\hline
1 & 1 & 1\\
1 & 0 & 1\\
1 & 1 & 1\\
\hline
\end{tabular}
\end{table}

Due specifically to our use of ReLU activation functions throughout our networks (rather than sigmoids), several copies of this filter must be applied in order to detect different specific neighbor counts. In particular, because the Game of Life rules require specific information about whether the total number of "alive" neighbors is $<2$, $2$, $3$, or $\geq4$, we need four duplicates of the neighbor counting filter, with biases $( -1, -2, -3, -4)$, in order to produce unique activation patterns for each neighbor total after the ReLU activation is applied.

We thus perform a single convolution of an $L\times L$ binary input image with $5$ total $3\times3\times1$ convolutional filters, producing an $L\times L \times 5$ activation volume.  Hereafter, we assume that the identity filter is the lowest-indexed filter in the stack, followed by the filters that count the successively-increasing numbers of neighbors $<2$, $=2$, $=3$, and $\geq4$.

Each $5 \times 1$ pixel across the $L \times L$ face of the activation volume now contains a unique activation pattern that can be matched against the appropriate output case. In the next layer of the network, two $1\times1$ convolutional filters with depth $5$ are applied
\[
(0, 0, 4/3, -8/3, -1/3)
\]
\[
(3/2, 5/4, -5, -1/4, -1/4)
\]
which are combined with biases $-1/3,-7/4$ and then activated with ReLU activation, resulting in an $L\times L \times 2$ activation volume. In order to generate a final $L\times L$ output corresponding to the next state of the automaton, this volume is summed along its depth---which can be performed efficiently as a final convolution with a $1\times1$ filter with value $(1,1)$ along its depth, and no bias. This will produce an $L\times L$ output image correspond to the next state of the Game.

For an example implementation of this algorithm in TensorFlow, see the function 
\begin{verbatim} 
ca_funcs.make_game_of_life()
\end{verbatim} in \url{https://github.com/williamgilpin/convoca/blob/master/ca_funcs.py}.

In principle, this architecture can work for any outer-totalistic cellular automaton, such as Life without Death, High Life, etc---although depending on the number of unique neighbor count and center pixel pairings that determine the ruleset, the number of neighbor filters may beed to be adjusted. For example, in the Game of Life the cases of $0$ living and $1$ living neighbors do not need to be distinguished by the network, because both cases result in the center pixel having a value of zero in the next timestep.

Likewise, for a purely totalistic cellular automaton (such as a majority vote rule), only a single convolutional filter (consisting of $9$ identical values) is necessary, because the value of the center pixel does not need to be resolved by the network.

\subsection{Neural network training details}

Convolutional neural networks were implemented in Python 3.4 using TensorFlow 1.8 \cite{abadi2016}. Source code is available at \url{https://github.com/williamgilpin/convoca}.

For all convolutions, periodic boundary conditions were implemented by manually copying pixel values from each edge of the input image, and then appending them onto the opposite edges. The padding option "VALID" was then used for the first convolutional filter layer in the TensorFlow graph.

Hyperparameters for the large networks described in the main text were optimized using a grid search. For each training run performed while optimizing hyperparameters, a new validation set of unseen binary images associated with an unseen cellular automaton ruleset was created, in order to prevent the cellular automaton ruleset from biasing the choice of hyperparameters. Once hyperparameters were chosen, and training on arbitrary cellular automata started, an additional validation set of binary images was generated for each ruleset. These images were used to determine when to stop training. Finally, an unseen set of binary images was used as a test partition, in order to compute the final accuracy of the trained networks. The training and test accuracies (before rounding the CNN output to the nearest integer) were within $0.3\%$ for all networks studied, which is a direct consequence of the network's ability to represent all input cases exactly. After rounding the CNN output to the nearest integer, both the train and test datasets had $100\%$ accuracy. The unrounded train and test performance during the training of one network are shown as a function of training epoch in Figure S1 of the supplementary material.

The default networks contained one $3\times3$ convolutional layer followed by $11$ layers of $1\times1$ convolutions. The convolutional layer, as well as the $1\times1$ layers, each had $100$ filters. A depth of $12$ layers was chosen for the network ensembles analyzed in the main text, in order to facilitate analysis of hidden layers across a variety of depths. Network and training parameters are given in Table \ref{params}.

\begin{table}[htp]
\centering
\caption{Hyperparameters for networks used in the main text.}
\vspace{4 mm}
\label{params}
\begin{tabular} {l|l}
{Parameter}     &  {  Value}	\\
\hline
\sf Input dimensions		&  \sf $10\times10$ px\\
\sf Number of layers		&  \sf $12$\\
\sf Neurons per layer		&  \sf $100$\\
\sf Input samples	&  \sf $500$ images\\
\sf Batch size	&  \sf $10$ images\\
\sf Weight initialization	 & He Normal\cite{he2015delving} \\
\sf Weight scale	& \sf $1$ \\
\sf Learning rate	& \sf $10^{-4}$ \\
\sf Max train epochs	& \sf $1500$\\
\sf Optimizer	& \sf \text{Adam}\\
\sf Loss	& \sf L2\\
\end{tabular}
\end{table}

We also considered the degree to which the exact dimensions of the "large network" affect our results. We trained another ensemble of networks with loss function, hyperparameters, and optimizer identical to the main text, but with the number of layers and the number of neurons per layer doubled (Table \ref{params_big}). As we observe in the main text, our results remain almost identical (Figure S2 of the supplementary material, left panel). We attribute this to the relatively small number of unique input cases that the networks need to learn ($512$) as compared to the potential expressivity of large networks. 

\begin{table}[htp]
\centering
\caption{Hyperparameters for the large network.}
\vspace{4 mm}
\label{params_big}
\begin{tabular} {l|l}
{Parameter}     &  {  Value}	\\
\hline
\sf Input dimensions		&  \sf $10\times10$ px\\
\sf Number of layers		&  \sf $24$\\
\sf Neurons per layer		&  \sf $200$\\
\sf Input samples	&  \sf $500$ images\\
\sf Batch size	&  \sf $10$ images\\
\sf Weight initialization	 & He Normal\cite{he2015delving} \\
\sf Weight scale	& \sf $1$ \\
\sf Learning rate	& \sf $10^{-4}$ \\
\sf Max train epochs	& \sf $1500$\\
\sf Optimizer	& \sf \text{Adam}\\
\sf Loss	& \sf L2\\
\end{tabular}
\end{table}

As a control against the choice of optimizer and loss affecting training, we also trained a replicate ensemble of networks that had the same network shapes ($12$ layers with $100$ neurons each) but a different loss function and optimizer, for which different optimal hyperparameters were found using a new grid search (Table \ref{params_alt}). We compare results using this alternative network to the default network described in the main text, and we find the results are nearly identical.

\begin{table}[htp]
\centering
\caption{Hyperparameters for the alternative network.}
\vspace{4 mm}
\label{params_alt}
\begin{tabular} {l|l}
{Parameter}     &  {  Value}	\\
\hline
\sf Input dimensions		&  \sf $10\times10$ px\\
\sf Number of layers		&  \sf $12$\\
\sf Neurons per layer		&  \sf $100$\\
\sf Input samples	&  \sf $500$ images\\
\sf Batch size	&  \sf $20$ images\\
\sf Weight initialization	 & He Normal\cite{he2015delving} \\
\sf Weight scale	& \sf $5\times10^{-1}$ \\
\sf Learning rate	& \sf $5\times 10^{-4}$ \\
\sf Max train epochs	& \sf $3000$\\
\sf Optimizer	& \sf S. G. D.\\
\sf Loss 	& \sf cross-entropy\\
\end{tabular}
\end{table}

Figures S2 (right panel) and S3 of the supplementary material show the results of training a network using these parameters. The shape of the training curve is slightly different, with the universal transient (during which the network learns general features of the input data such as the range and number of unique cases) being much longer for this network. However, the later phases of training continue similarly to the standard network, with $\mathcal H_{ca}$ strongly affecting the later stages of training and the final loss. Moreover, after training has concluded, the dependence of the internal representations of the network on $\mathcal H_{ca}$ (Figure S2 of the supplementary material) matches the patterns seen in the default network above.

\section{Supplementary Materials and Additional Experiments}

For this arXiv submission, additional supplementary material has been uploaded as an ancillary file. To obtain the SI, navigate to the abstract landing page, and then on the right-hand side select "Other formats" under the "Download" header. On the resulting page, click the link for "Download source" under the "Source" header. A numbered file will download. Either extract the file from the terminal, rename the file with the extension ".gz", and then double-click the file to extract a directory containing the PDF of supplementary material.

\bibliography{ca_cites}
\end{document}